\documentclass[useAMS,usenatbib,usegraphicx]{mn2e}
\usepackage{color}
\usepackage{graphicx}
\usepackage{amssymb}

\title[SAX J1324.4--6200 and SAX J1452.8--5949]{{\it Chandra} and {\it XMM-Newton} 
observations of the low-luminosity X-ray pulsators SAX J1324.4--6200 and SAX J1452.8--5949}
\author[Kaur et al.]{Ramanpreet Kaur$^{1}$\thanks{E-mail: raman@aries.ernet.in}, 
Rudy Wijnands$^{2}$,
Alessandro Patruno$^{2}$, 
Vincenzo Testa$^{3}$, 
\newauthor 
GianLuca Israel$^{3}$,
Nathalie Degenaar$^{2}$,
Biswajit Paul$^4$,  
Brijesh Kumar$^1$ \\ 
$^1$ Aryabhatta Research Institute of observational sciences, 
Manora Peak, Naini Tal, 263\,129, India\\
$^2$ Astronomical Institute "Anton Pannekoek", University of Amsterdam, 
Kruislaan 403, 1098 SJ, Amsterdam, The Netherlands \\
$^3$ INAF - Osservatorio Astronomico di Roma, via Frascati 33, 00040 
Monte Porzio Catone, Italy \\ 
$^4$ Raman Research Institute, C. V. Raman Avenue, Sadashivanagar, Bangalore 560 080, India.\\}
\begin{document}

\date{}

\pubyear{2008}

\maketitle

\label{firstpage}

\begin{abstract}
We present results from our {\it Chandra} and {\it XMM-Newton} observations 
of two low-luminosity X-ray pulsators SAX J1324.4--6200 and SAX J1452.8--5949 which
have spin-periods of 172 s and 437 s respectively. The 
{\it XMM-Newton} spectra for both sources can be fitted well with a simple 
power-law model of photon index $\sim$ 1.0. A black-body model can equally well 
fit the spectra with a temperature of $\sim$ 2 keV for both sources. During our 
{\it XMM-Newton} observations, SAX J1324.4--6200 is detected with coherent 
X-ray pulsations at a period of $172.86 \pm 0.02$ s while no  
pulsations with a pulse fraction greater than 15 \% (at 98\% confidence level) 
are detected in SAX J1452.8--5949. The spin period 
of SAX J1324.4--6200 is found to be increasing on a time-scale of $\dot{P}$ 
= $(6.34 \pm 0.08) \times 10^{-9}$ s s$^{-1}$ which would suggest that the
accretor is a neutron star and not a white dwarf. Using sub-arcsec spatial 
resolution of the {\it Chandra} telescope, possible counterparts are seen for 
both sources in the near-infrared images obtained with the SOFI instrument on the {\it New 
Technology Telescope}. The X-ray and near-infrared properties of SAX J1324.4--6200 
suggest it to be either a persistent high mass accreting X-ray pulsar or  
a symbiotic X-ray binary pulsar at a distance $\le$ 9 kpc. We identify the infrared counterpart
of SAX J1452.8--5949 to be a late-type main sequence star at a distance $\le$ 10 kpc,
thus ruling out SAX J1452.8--5949 to be a high mass X-ray binary. However with the 
present X-ray and near-infrared observations, we cannot make any further conclusive 
conclusion about the nature of SAX J1452.8--5949. 
\end{abstract}


\section{Introduction}

In the past ten years, using the Advanced Satellite for Cosmology 
and Astrophysics ({\it ASCA}), {\it BeppoSAX} and the Rossi X-ray 
Timing Explorer ({\it RXTE}) satellites, a population of faint 
(L$_X \lesssim 10^{36}$ erg s$^{-1}$) X-ray sources has been 
found in our Galaxy which harbor a slow pulsating source with 
periods ranging from several seconds to over a thousand seconds. 
These `slow' pulsators are found to harbor a variety of source types
like anomalous X-ray pulsars (isolated slowly rotating neutron stars; e.g.
Torii et al. 1998), accreting magnetized white dwarfs (i.e. intermediate
polars, AM Her type systems; e.g., Misaki et al. 1996), and neutron stars
accreting from a high-mass companion star (i.e., mostly as Be/X-ray transients; 
e.g., Hulleman et al. 1998). Despite the successes in determining the 
nature of these slow pulsators, there remains a group of persistent 
systems  whose nature still has not been determined. The X-ray properties 
of these sources suggest that most of them are neutron stars accreting
from a high mass companion star, however accreting white dwarf cannot
be excluded. Furthermore, in some cases, the possibility of a system in which 
a neutron star accretes from a low-mass companion star also cannot be 
excluded (Lin et al. 2002). More observations at all wavelengths 
(i.e., X-ray or near IR) could help to unveil the nature of these pulsators.
In this paper, we present the results from our {\it Chandra}, {\it XMM-Newton} 
and {\it New Technology Telescope} (NTT) observations of two faint X-ray pulsators 
SAX J1324.4--6200 ($l$ = 306$^{\circ}.79$, $b$ = 0$^{\circ}$.60) 
and SAX J1452.8--5949 ($l$ = 317$^{\circ}.65$, $b$ = --0$^{\circ}.46$). 

SAX J1324.4--6200 (hereafter SAX13), which has a pulse period 
of $\sim$ 170 s, was discovered serendipitously from observations of the low 
mass X-ray binary (LMXB) XB 1323--619 performed on August 22, 1997 
with {\it BeppoSAX} (Angelini et al. 1998). The X-ray (1.8--10 keV) 
spectrum of the source could be fitted with either an absorbed 
power-law with photon index of 1.0 $\pm$ 0.4 and hydrogen column 
density, $N_{\mathrm{H}}$ of $7.8^{+2.7}_{-1.1} 
\times 10^{22}$ cm$^{-2}$ or with a black-body model with a 
temperature, kT of 2.4 $\pm$ 0.4 keV and $N_{\mathrm{H}}$ of 4$^{+3}_{-2}$ 
$\times$ 10$^{22}$ cm$^{-2}$ (Angelini et al. 1998). SAX13 
was also detected in the observations performed on XB 1323--619 
using {\it ASCA} on August 4, 1994 (Angelini et al. 1998). In addition, 
pointed {\it ASCA} observations (of 187 ks) were performed on SAX13 
on February 2, 2000 and the source was detected with a pulse 
period of $\sim$ 171.2 s (Lin et al. 2002). Lin et al. (2002) 
found a possible orbital period of the system of 27 $\pm$ 1 hour and 
suggested that the system could be a LMXB pulsar. Recently, during 
short observations performed using {\it Swift} on December 30, 
2007, SAX13 was detected  with a pulse period of 172.8 s 
(Mereghetti et al. 2008) which would imply a spin down over 
the last 10 years of  $\dot{P}$  = $\sim$ 6 $\times$ 10$^{-9}$ 
s s$^{-1}$. Mereghetti et al. (2008) identified a possible 2MASS
near-infrared counterpart in the {\it Swift} error circle of SAX13  
with a {\it K} band magnitude of 14.39 $\pm$ 0.08 and suggested 
that the source could be a persistent Be accreting X-ray pulsar.
 
SAX J1452.8--5949 (hereafter SAX14) was discovered 
using {\it BeppoSAX} with a spin period of 
$\sim$ 437 s (Oosterbroek et al. 1999). The X-ray spectra 
of the source could be fitted well with an absorbed power-law 
model with a photon index of 1.4 $\pm$ 0.6. Oosterbroek et al. (1999) suggested
that SAX14 could be an accreting Be X-ray pulsar at a distance 
of $6 -12$ kpc with the luminosity  of $\sim 10^{34}$ erg s$^{-1}$.


\begin{table*} \label{tab:observation}
\centering
\caption{Log of the X-ray observations of SAX J1324.4--6200 and SAX J1452.8--5949.}
\label{observations_log}
\begin{tabular}{|ccccccc|}
\hline
Object & Telescope/Instrument & Date & ObsId & Total Observation     \\
      &           &   (UT)   &       &  Span     (ks)          \\
\hline
SAX J1324.4--6200&{\it Chandra}/ACIS-I  &   28 Nov 2007  &  9012      &  1.1  \\
SAX J1452.8--5949 &{\it Chandra}/ACIS-I  &   30 Dec 2007  &  9014       &  1.0  \\
SAX J1324.4--6200&{\it XMM-Newton}/EPIC &   11 Jan 2008  & 0511010201  & 19.3\\
SAX J1452.8--5949 &{\it XMM-Newton}/EPIC &   07 Feb 2008  & 0511010501   &  6.9 \\
\hline
\end{tabular}
\end{table*}


\section{X-ray Observations}

Our X-ray observations of SAX13 and SAX14 were carried out using the 
European Photon Imaging Camera (EPIC) aboard the
{\it XMM-Newton} satellite and advanced CCD Imaging Spectrometer 
(ACIS) aboard the {\it Chandra} satellite.

\subsection{\it XMM-Newton} 

SAX13 and SAX14 were observed for $\sim$ 19 ks on January 11, 2008 
(Obs ID. 0511010201) and $\sim$ 7 ks on February 7, 2008 (Obs ID. 
0511010501) respectively. Both the EPIC-{\it MOS} and EPIC-{\it pn} 
cameras (Turner et al. 2001; Struder et al. 2001) were operated in the 
{\it Full Frame} mode and with the medium filter. The observation 
details are summarized in Table \ref{observations_log}. The EPIC 
observation data files were processed using the {\it XMM}-Science 
Analysis System {(SAS version 7.1.0)\footnote[1]
{See http://xmm2.esac.esa.int/sas/}}. 
Investigation of the full-field count-rate of SAX13 revealed $\sim$ 
3 ks of high-count rate background particle flaring and is removed 
from the data. However we 
did not see any significant background flaring in SAX14 observation.
We used the task {\it edetect\_chain} to find the exact position of  
X-ray source in the combined images of EPIC-{\it MOS} and EPIC-{\it 
pn} instruments. In the {\it BeppoSAX} error circle of SAX13 
(Angelini et al. 1998), we detected only one source in the 
{\it XMM-Newton} observations at a position of R.A. = 13$^\mathrm{h}$ 
24$^\mathrm{m}$ 26$\fs$64 and DEC. = --62$^{\circ}$ 01$^{\prime}$ 
18${\farcs}$48 with an error circle of radius 2${\farcs}$0 (90\% confidence; 
J2000; all coordinates in the paper are for J2000 epoch). 
In the {\it BeppoSAX} error circle of SAX14 (Oosterbroek et al. 1999), only 
one source is detected in the {\it XMM-Newton} observations at a 
position of R.A. = 14$^\mathrm{h}$ 52$^\mathrm{m}$ 
52$\fs$80 and DEC. = --59$^{\circ}$ 49$^{\prime}$ 08${\farcs}$04 with 
an error circle of radius 2${\farcs}$0 (90\% confidence). The error circle on 
the position of X-ray source is adopted as a quadratic sum of the bore 
sight error of the {\it XMM-Newton} {telescope\footnote[2] 
{See http://xmm2.esac.esa.int/docs/documents/CAL-TN-0018.pdf} 
and the statistical error given by the task {\it edetect\_chain}.

\subsection{\it Chandra}

SAX13 and SAX14 were observed for $\sim$ 1 ks each on November 28, 2007 
(Obs ID. 9012)  and December 30, 2007 (Obs ID. 9014) respectively. The
details of the observations are given in Table \ref{observations_log}. 
The data was obtained with the ACIS-I CCDs operating in the FAINT mode
for both sources. We processed the event 2 files using the standard 
software packages CIAO {4.0\footnote[3] {{\it Chandra} Interactive Analysis 
of Observations (CIAO), http://cxc.harvard.edu/ciao/.}} and CALDB 
{3.4.2\footnote[4] {{\it Chandra} Calibration Database (CALDB), 
http://cxc.harvard.edu/caldb/}. The task {\it wavdetect} was used to 
find the exact position of X-ray source in the image. One source is 
detected in the {\it XMM-Newton} error circle of SAX13 with a total 
of 87 counts at a position of R.A. = 13$^\mathrm{h}$ 24$^\mathrm{m}$ 
26$\fs$70 and DEC. = -62$^{\circ}$ 01$^{\prime}$ 19${\farcs}$49 with 
the error circle of radius 0${\farcs}$65 (Figure \ref{fig:sax13_chart}). In the
{\it Chandra} image of SAX14, one source is detected (Figure \ref{fig:sax14_chart})
in the {\it XMM-Newton} error circle of SAX14, albeit with only 4 counts 
at a position of R.A. = 14$^\mathrm{h}$ 52$^\mathrm{m}$ 52$\fs$70 and DEC. = -59$^{\circ}$ 
49$^{\prime}$ 8${\farcs}$07 with the error circle of radius 0${\farcs}$95. 
The average background count is $\sim$ 1 count  
for a similar area in the full
{\it Chandra} image of SAX14, indicates that indeed we have detected 
SAX14. The errors on the positions of SAX13 and SAX14 are obtained as 
a quadratic sum of the bore sight error of the {\it Chandra} 
{telescope\footnote[5] {See http://cxc.harvard.edu/cal/ASPECT/celmon.}, 
1-$\sigma$ {\it wavedetect} errors and a contribution that depends on 
the number of detected counts (van den Berg et al. 2004; especially 
important for SAX14).

\section{Infrared Observations and Data Analysis}

Near-infrared observations in the {\it J, H, K$_{\mathrm{s}}$} wavebands 
were performed in February 2001 with the 3.58 m ESO-NTT  
telescope equipped with the near-infrared imager and spectrograph SOFI.
The instrument was set up in large imaging mode with a pixel 
scale of 0${\farcs}$29 and a field of view of $\mathrm{5}^{\prime}.0\times 
\mathrm{5}^{\prime}.0$ when SAX13 was observed and in small imaging mode
with a pixel scale of 0${\farcs}$14 and a field of view of
$\mathrm{2}^{\prime}.5 \times \mathrm{2}^{\prime}.5$ when 
SAX14 was observed (both in double `correlated' readout mode). 
In both cases, the seeing conditions were very good, ranging 
from $0{\farcs}58$ to $0{\farcs}66$ between $K_{\mathrm{s}}$ and {\it J} bands.
Images were acquired with the auto-jitter sequence, obtaining
several dithered frames for each filter, that were then re-aligned
and co-added together. Table \ref{tab:observations}  summarizes 
the observations. The acquisition mode is the usual one for 
near-IR arrays: a number of single frames (NDIT) having exposure 
times of DIT (Detector Integrator Time) seconds are acquired 
and then co-added aboard generating an output image having exposure 
time equal to one DIT, but the signal-to-noise (S/N) statistics of NDIT$\times$DIT seconds.
All the output frames (Nframes) are then processed and co-added
offline, generating the final science image.

\begin{table*}     
\caption{Log of the 3.58 m ESO-NTT near-infrared observations of 
SAX J1324.4--6200 and SAX J1452.8--5949. NDIT represents the 
Number of single frames, having exposure times of DIT 
(Detector Integrator Time) seconds and are used to generate 
an output image having exposure time equal to one DIT. 
The number of output frames are represented
by Nframes.} 
\centering
{\scriptsize  
\begin{tabular}{cccccccccccc} 
\hline 
\hline   
Source & Date & \multicolumn{3}{c}{{\it J}} & \multicolumn{3}{c}{{\it H}} & \multicolumn{3}{c}{{\it K}$_{\mathrm{s}}$} & FWHM(arcsec)\\
       & (UT) &  DIT & NDIT & Nframes & DIT & NDIT & Nframes & DIT & NDIT & Nframes & \\   
\hline   
SAXJ1324.4--6200 & 06 Feb 2001 & 3 & 5 & 8 & 3 & 5 & 8 & 3 & 5 & 12 & 0.6\\ 
SAXJ1452.8--5949 & 08 Feb 2001 & 3 & 5 & 10 & 3 & 5 & 12 & 3 & 5 & 16 & 0.6\\  
 \hline
\end{tabular}}  
\label{tab:observations}  
\end{table*}                       

Images were reduced using the package {\it eclipse} developed by ESO 
to handle  near-IR observations. The pipeline executes the following 
steps: first, a master 'sky' images is obtained by median stacking
all the frames for one observation and for each filter. This 'sky' 
image is subtracted to every single image generating sky-subtracted 
frames. This step also subtracts bias and dark current contributions. 
The images are then flat-fielded with flat field images obtaining on 
the dome. In the case of SOFI, also a 'screen' correction image applied 
to correct for residual illumination gradients in the dome. At the end, 
the pre-reduced images are aligned and average stacked, with a proper
$\sigma$-clipping selection, obtaining the science-ready final images.
Source detection and photometry has been performed with the {\it daophot} 
package within IRAF\footnote{Image Reduction and Analysis Facility 
(IRAF) is distributed by the National Optical 
Astronomy Observatories, which are operated by the Association of 
Universities  for Research  in Astronomy,  Inc., under cooperative 
agreement with  the National Science Foundation.}. In particular, 
a second-order variable PSF model has been computed for our images 
to minimize positional biases due to the (small) PSF variation over the field.
The output magnitudes have been aperture corrected and calibrated with the 
2MASS catalog (Skrutskie et al. 2006). The same 2MASS has been used 
also to obtain an astrometric calibrations of the frames. The interactive
software Skycat-GAIA has been used, selecting the portion of catalog of 
the area and fitting a solution to the image. The positions obtained 
in this way have (absolute) uncertainties of about 0${\farcs}$2, 
and relative uncertainties far below a fraction of a pixel.


\begin{figure*}
\begin{tabular}{c}
\resizebox{13cm}{!}{\includegraphics{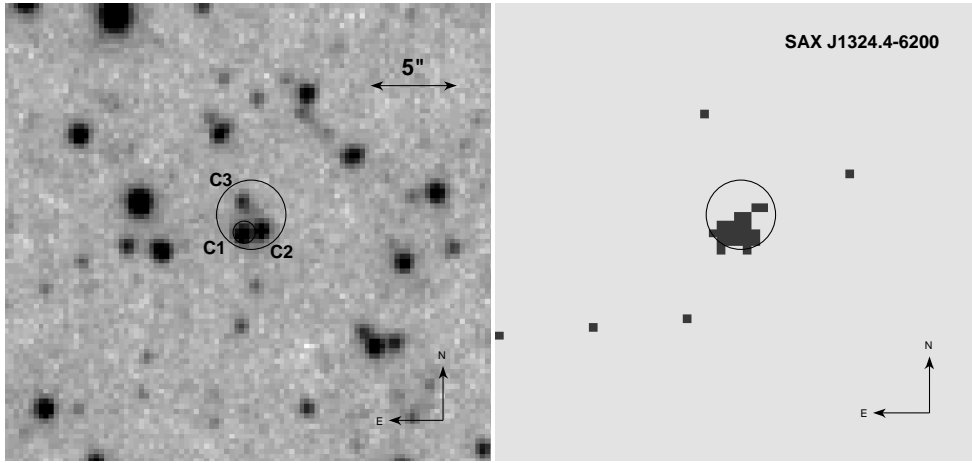}}
\end{tabular}
\caption{{\it Left} : The {\it K$_{\mathrm{s}}$} waveband image of SAX J1324.4--6200
taken by SOFI instrument on the New Technology Telescope. The {\it XMM-Newton} error 
circle of radius 2$\arcsec$ and the {\it Chandra} error circle of radius 
0.65$\arcsec$ of SAX J1324.4--6200 are also plotted. {\it Right} : 
{\it Chandra} ACIS-I image of SAX J1324.4--6200 with the {\it XMM-Newton} error circle of 
radius 2$\arcsec$.}
\label{fig:sax13_chart}
\end{figure*}

\begin{figure*}
\begin{tabular}{c}
\resizebox{13cm}{!}{\includegraphics{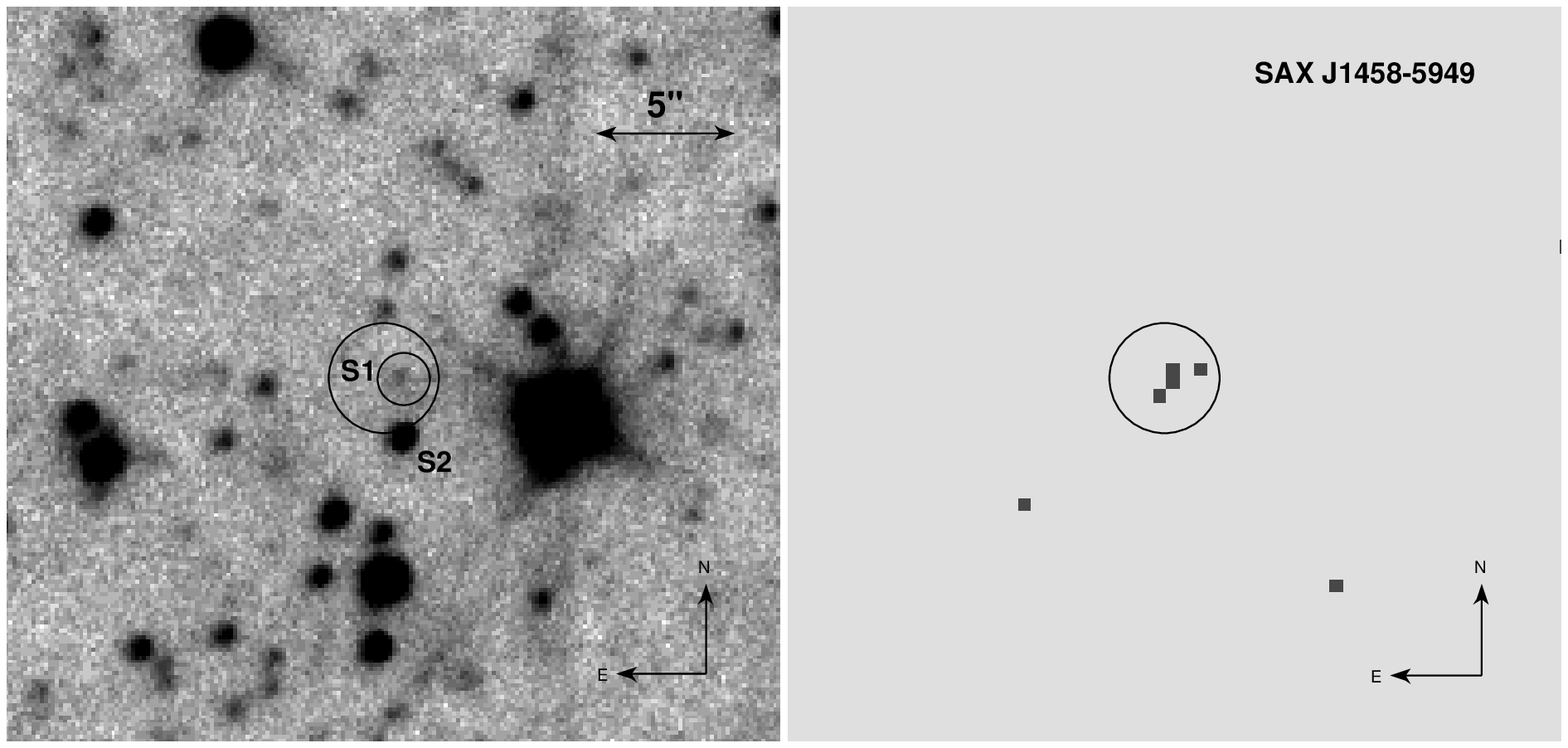}}
\end{tabular}
\caption{{\it Left} : The {\it K$_{\mathrm{s}}$} waveband image of SAX J1452.8--5949 
taken by SOFI instrument on the New Technology Telescope. The {\it XMM-Newton} 
error circle of radius 2$\arcsec$ and the {\it Chandra} error circle of radius 
0.65$\arcsec$ of SAX J1452.8--5949 are also plotted.
{\it Right} : {\it Chandra} ACIS-I image of SAX J1452.8--5949 with the {\it XMM-Newton}
error circle of radius 2$\arcsec$.}
\label{fig:sax14_chart}
\end{figure*}

As can be seen in the $K_{\mathrm{s}}$ image taken with NTT, the position of three point 
sources labeled as `C1', `C2', `C3' are consistent with the {\it XMM-Newton} 
error circle of SAX13 (Figure \ref{fig:sax13_chart}). The position of two point 
sources labeled as `S1' and `S2' are consistent with the {\it XMM-Newton} 
error circle of SAX14 (Figure \ref{fig:sax14_chart}). The {\it Chandra} error
circle for both the sources is also shown : the position of star `C1' falls in the
{\it Chandra} error circle of SAX13 and that of `S1' in the {\it Chandra} error circle of SAX14. 
Therefore we consider `C1' as the near-infrared counterpart of SAX13 and `S1' of  
SAX14. We have listed the positions 
(in Right Ascension and Declination) and magnitudes in {\it J, H, K$_{\mathrm{s}}$} bands for the 
near-infrared counterparts of SAX13 and SAX14 in Table \ref{mag}. For the sake of 
completeness, the positions and magnitudes of the other stars falling in the 
{\it XMM-Newton} error circles of SAX13 and SAX14 are also listed in Table \ref{mag}.


\begin{table*} \label{tab:mag}
\centering
\caption{{\it J, H and K$_{\mathrm{s}}$} magnitudes of stars falling in the {\it XMM-Newton} error circle
of SAX J1324.4--6200 and SAX J1452.8--5949. The stars listed here are also marked in
the near-infrared ESO-NTT images shown in Figure \ref{fig:sax13_chart} and \ref{fig:sax14_chart}.}
\label{mag}
\begin{tabular}{|cccccc|}
\hline
Star &  R.A. & DEC.   & J    & H    &   $K_{\mathrm{s}}$  \\ 
 &  hh:mm:ss &  $^{\circ}$ $^{\prime}$ ${\arcsec}$  & magnitude      & magnitude   &   magnitude  \\ \hline
   & \multicolumn{4}{c}{Stars in {\it XMM--Newton} error circle of SAX J1324.4--6200 }  & \\ \hline
C1 & 13:24:26.71 & --62:01:19.59 & 19.57 $\pm$ 0.11 & 16.61 $\pm$ 0.09 & 14.97 $\pm$ 0.11  \\
C2 & 13:24:26.56 & --62:02:19.38 & 18.76 $\pm$ 0.10 & 16.48 $\pm$ 0.09 & 15.45 $\pm$ 0.11  \\
C3 & 13:24:26.72 & --62:01:17.71 & 18.37 $\pm$ 0.10 & 17.00 $\pm$ 0.09 &  16.62 $\pm$ 0.12 \\ \hline
   & \multicolumn{4}{c}{ Stars in {\it XMM--Newton} error circle of SAX J1452.8--5949}  &   \\ \hline

S1 & 14:52:52.72 & --59:49:07.92 &  18.59 $\pm$ 0.12 & 17.75 $\pm$ 0.12 & 17.93 $\pm$ 0.12 \\
S2 & 14:52:52.70 & --59:49:10.14 &  16.92 $\pm$ 0.11 & 15.50 $\pm$ 0.10 & 15.06 $\pm$ 0.10 \\
\hline
\end{tabular}
\end{table*}


\section{Data Analysis and Results}

\subsection{Timing Analysis} 

The {\it XMM-Newton} EPIC-{\it pn} data were used for the timing analysis 
for both SAX13 and 
SAX14. The X-ray events were extracted in a circular region of 
radius 15$\arcsec$ and 30$\arcsec$ for SAX13 and SAX14, respectively, 
centered on the position of the target in the {EPIC-{\it pn} image. 
We were forced to take smaller circular region for SAX13 as it was 
located close to a CCD gap. The background X-ray events were extracted 
with a similar circular region on the same CCD in a source-free region. 
The times of the events were transformed to barycentric times using 
the {\it SAS} tool {\it barycor} using the {\it Chandra} position of the source 
and the JPL-DE405 ephemeris. The events were re-binned with a time resolution 
of 0.2 s and the light curves were corrected for background contamination. 

We reduced $\approx 16$ ks of pointed observations from the 
EPIC-{\it pn} CCD camera for SAX13 and detected the source with a 
background corrected count rate of 0.28 counts s$^{-1}$. We 
folded the background corrected light curve in five chunks of  
$\approx 3000$ s, with the best spin-period available from Mereghetti et al. (2008). 
The pulse profiles were then cross correlated with a template profile
obtained by folding the entire light curve. The cross correlation returns
the time of arrivals (TOAs) of each pulse profile with the fiducial
point for the measurement fixed to be the peak of the template
profile. The technique applied to measure the TOAs and their
statistical uncertainties closely resemble the radio pulsar technique
(e.g., Taylor 1992). The TOAs were then phase connected with a
linear polynomial fit of the pulse phase to obtain the pulsar spin period. 
The measure of a spin period
derivative using a phase coherent technique is prevented given the
short baseline of the observation for which the phase-connected
solution could be applied. Our measured spin period is
$P_{\mathrm{s}}=172.86\pm0.02$ s (referred at MJD 54460; The error is calculated
with 68\% confidence level) and is consistent with
the previous measurements of Mereghetti et al. (2008) that reported a spin
period of $172.84 \pm 0.1$ s at MJD 54464.2. The phase-connected solution
has not the precision required for an extrapolation to the
previous observations of the source (see Table \ref{p_observation}). Therefore we
fitted all the previous and our new measured value of the spin period
(Angelini et al. 1998, Lin et al. 2002, Mereghetti et al. 2008) 
with a linear relation $P_{\mathrm{s}}(t)=P_{\mathrm{0}} + \dot{P}_{\mathrm{s}}t$,
where $\dot{P}_{\mathrm{s}}$
is the spin period derivative of the pulsar and $P_{\mathrm{0}}$ is the 
spin period at the time $t=0$.  The
value obtained from our fit is $\dot{P_{\mathrm{s}}}=(6.34 \pm 0.08)\times
10^{-9}$ s s$^{-1}$ with a $\chi^{2}$ of 4.46 for 3 degrees of
freedom. The background subtracted pulse profile of SAX13 in the 
energy range 0.2 - 12 keV is shown in Figure \ref{pulseprofile}.


\begin{figure}
\begin{tabular}{c}
\resizebox{8.3cm}{!}{\includegraphics[ angle =-90]{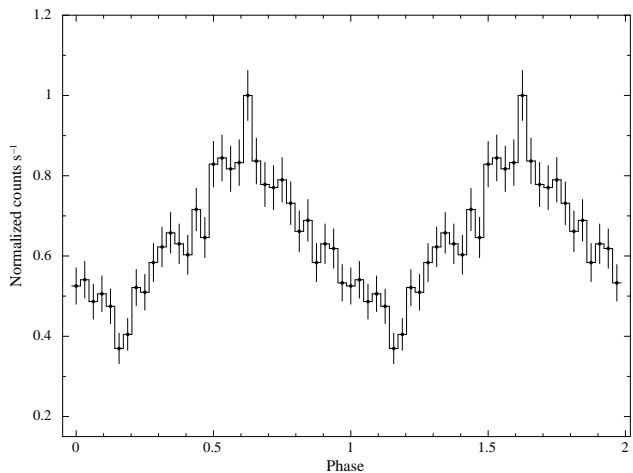}}
\end{tabular}
\caption{The {\it XMM-Newton} EPIC-{\it pn} 0.2--12 keV background subtracted
pulse profile of SAX J1324.4--6200. Two cycles are shown for clarity.}
\label{pulseprofile}
\end{figure}


We also folded the entire {\it XMM-Newton} light curve in one single profile
to increase the S/N and measure the harmonic content and
the fractional amplitude of the pulsation. The pulse profile is fitted
well with a single sinusoid with a significance larger than
20$\sigma$ and has a fractional amplitude of $(52 \pm 4)\%$.

SAX14 was detected with a background subtracted count rate of 0.05 counts s$^{-1}$. 
We repeated the same procedure (described above) for SAX14 using the best available 
spin period (Oosterbroek et al. 1999). However no significant pulse profiles have 
been detected in the energy band 0.2 - 12 keV with an upper limit on the pulsed 
fraction amplitude (Vaughan et al. 1994) of 15 \% (at 98\% confidence level). 
To further investigate the presence of pulsations, we 
divided the EPIC-{\it pn} light curve into soft (0.2 - 4 keV) and hard (4 - 12 keV) 
energy bands with equal count rate. No significant pulse profiles were detected in both soft 
and hard energy bands with an upper limit on the pulse fractional amplitude of 
18\% and 22\% (at 98\% confidence level) respectively. 


{
\begin{table*}
\centering
\caption{Spectral Parameters for SAX J1324.4-6200 and SAX J1452.8--5949 for {\it XMM-Newton} EPIC observations. }
\label{parameters}
\begin{tabular}{lccc} \hline

                                                    & \multicolumn{1}{c}{{ SAX J1324.4-6200}}      & \multicolumn{2}{c}{{SAX J1452.8--5949}} \\ \hline
Parameter                                           & \multicolumn{2}{c}{Spectral parameters of model (Absorption + power-law)}  &     \\ \hline
N$_H$ x 10$^{22}$ (cm$^{-2}$)                       & 5.81    $\pm$ 0.65         &  1.22 $\pm$ 0.69     \\
Photon Index                                            & 1.01    $\pm$ 0.14         &  0.83 $\pm$ 0.28     \\
Reduced $\chi^2$/dof                                & 0.95/383                   &  0.6/19              \\
Observed Flux  2 - 10 keV (ergs cm$^{-2}$ s$^{-1}$)  & (4.5 $\pm$ 1.3) $\times$ 10$^{-12}$      & (5.4 $\pm$ 2.5) $\times$ 10$^{-13}$  \\
Unabsorbed Flux 2 - 10 keV (ergs cm$^{-2}$ s$^{-1}$) & (6.0 $\pm$ 1.4) $\times$ 10$^{-12}$      & (5.8 $\pm$ 2.9) $\times$ 10$^{-13}$  \\ \hline
Parameter                                           & \multicolumn{2}{c}{Spectral parameters of model (Absorption + black-body)}  &     \\ \hline
N$_H$ x 10$^{22}$ (cm$^{-2}$)                       & 3.07 $\pm$ 0.42            & $<$ 0.54 (90\% confidence) \\
Blackbody temperature (keV)                         & 2.24 $\pm$ 0.14            &  1.82 $\pm$ 0.24         \\
Reduced $\chi^2$/dof                                & 1.0/383                    &  0.7/20                 \\
Observed Flux 2 - 10 keV (ergs cm$^{-2}$ s$^{-1}$)    & (4.4 $\pm$ 0.2) $\times$ 10$^{-12}$      &  (4.9 $\pm$ 0.4)$\times$ 10$^{-13}$  \\
Unabsorbed Flux 2 - 10 keV  (ergs cm$^{-2}$ s$^{-1}$) & (5.0 $\pm$ 0.3) $\times$ 10$^{-12}$      &  (5.0 $\pm$ 0.4) $\times$ 10$^{-13}$ \\ \hline
\end{tabular}
\flushleft
\end{table*}
}

\subsection{Spectral Analysis}


\begin{figure}
\begin{tabular}{c}
\resizebox{8.3cm}{!}{\includegraphics[ angle =-90]{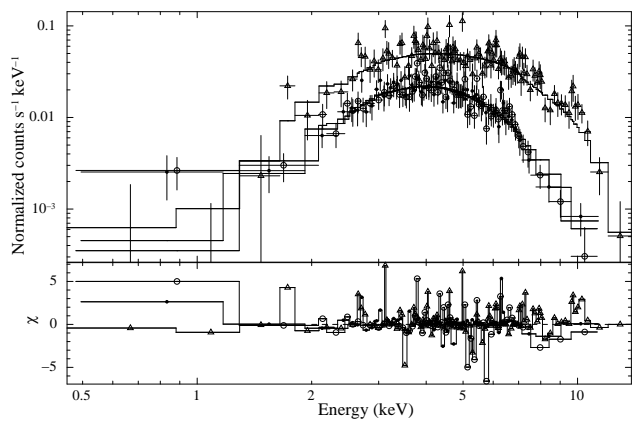}}
\end{tabular}
\caption{The {\it XMM-Newton} EPIC-{\it MOS} and {\it pn} spectra of SAX J1324.4--6200 fitted 
with an absorbed power-law model. The {\it pn} data points are represented by open triangles, 
the {\it MOS1} data points are represented by open circles and the {\it MOS2} data points 
are represented by filled circle.}
\label{sax13spectra}
\end{figure}


\begin{figure}
\begin{tabular}{c}
\resizebox{8.3cm}{!}{\includegraphics[ angle =-90]{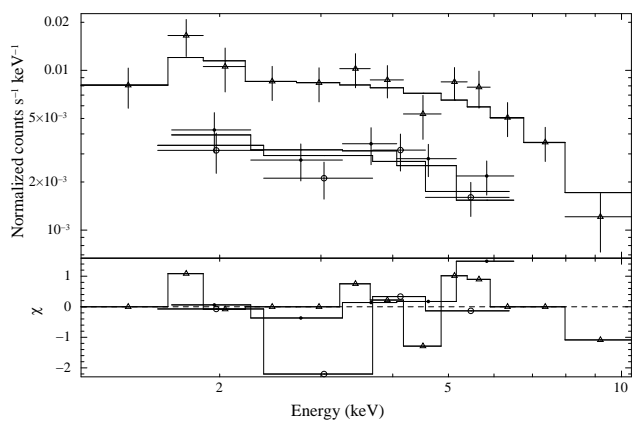}}
\end{tabular}
\caption{The {\it XMM-Newton} EPIC-{\it MOS} and {\it pn} spectra of SAX J1452.8--5949 fitted 
with an absorbed power-law model. The {\it pn} data points are represented by open triangles,
the {\it MOS1} data points are represented by open circles and {\it MOS2} are represented by filled circle.}
\label{sax14spectra}
\end{figure}


The {\it XMM-Newton} EPIC-{\it MOS} and {\it pn} data are used for spectral analysis
for SAX13 and SAX14. We used the same extraction region for EPIC-{\it pn} data as 
were used for the timing analysis reported in \S 4.1 while a 30$\arcsec$ circular region 
is used to extract the X-ray counts in the EPIC-{\it MOS} images for both sources.
SAS tool {\it xmmselect} was used to extract both source and background spectra. We used 
{\it XSPEC} (version 11.0.0) for our spectral analysis. The resulting spectra were 
re-binned to have minimum of 20 counts per bin. 

We fitted both power-law and black-body models to the SAX13 EPIC-{\it MOS} and 
{\it pn} X-ray spectra together. Both models could fit the data with a 
reduced $\chi^2$ of $\sim 1.0$ for 385 degrees of freedom (dof). The full 
values of obtained fit parameters are given in Table \ref{parameters}. For the 
power-law model, the spectrum could be fitted with a photon index of $1.0$ 
and an $N_{\mathrm{H}}$ of $5.8 \times 10^{22}$ cm$^{-2}$ while using a black-body model, 
we obtained a kT = 2.2 keV and an $N_{\mathrm{H}}$ of $3.1 \times 10^{22}$ cm$^{-2}$.
The observed flux for both models in the 2 - 10 keV energy band is $\sim$ 
$5 \times 10^{-12}$ erg s$^{-1}$ cm$^{-2}$.  With the present data, we are 
unable to distinguish between the power-law and black-body model fit. The power-law 
fit to SAX13 {\it XMM-Newton} EPIC-{\it MOS} and {\it pn} spectra is shown 
in Figure \ref{sax13spectra}. No significant Fe 6.4 keV emission line is seen 
in the combined {\it XMM-Newton} EPIC-{\it MOS} and {\it pn} spectrum of SAX13 
(Fig \ref{sax13spectra}). To find an upper limit on Fe 6.4 keV emission line,
we fixed the line-center at 6.4 keV in the spectrum and fitted a Gaussian to
the line. The upper limit on the equivalent width of the 6.4 keV emission line
is determined to be 118 eV (90\% confidence limit).

Similarly, we fitted both power-law and black-body models to the SAX14  
spectra. Both models provide acceptable fits with reduced $\chi^2$ in the 
range of 0.6 - 0.7 for 20 dof. From the power-law model, we obtained 
a photon index, $\Gamma$ of $0.8$ and an $N_{\mathrm{H}}$ of $1.2 \times 10^{22}$
cm$^{-2}$ (shown in Figure \ref{sax14spectra}) while using the black-body model   
we obtained a kT = 1.8 keV and an $N_{\mathrm{H}}$ = $0.5 \times 10^{22}$ cm$^{-2}$. 
The observed flux in 2 - 10 keV energy band (for both the models) is 
$\sim 5 \times 10^{-13}$ erg s$^{-1}$ cm$^{-2}$. The spectral parameters 
of SAX14 are listed in Table \ref{parameters}. No significant Fe 6.4 keV emission
line is seen in the SAX14 spectra (Fig \ref{sax14spectra}). However, an upper limit on iron 6.4 keV 
emission line in the combined {\it XMM-Newton} EPIC - {\it MOS} and {\it pn} 
spectrum of SAX14 is 436 eV (at 90\% confidence level).

\section{Discussion}

We carried out observations with the {\it Chandra} and {\it XMM-Newton} 
satellites to investigate the nature of the accreting 
X-ray pulsators SAX J1324.4--6200 and SAX J1452.8--5949. In addition, we 
obtained near-infrared data using the ESO-NTT to search for infrared counterparts 
for these pulsators. 

\subsection{SAX J1324.4--6200}

During our {\it XMM-Newton} observation, SAX13 is detected with a pulse 
period of $172.85 \pm 0.02$ s, consistent with the pulse period obtained 
by Mereghetti et al. (2008) from observations made by {\it Swift} on 
December 30, 2007, just 10 days before our {\it XMM-Newton} observations. 
The spin period history of SAX13 clearly shows a linear increase in pulse 
period from 170.35 to 172.86 s over the last 14 years (Table 
\ref{p_observation}) with spin period derivative  $\dot{P}$ = $(6.34 
\pm 0.08) \times 10^{-9}$ s s$^{-1}$. The spin period of high mass 
X-ray binary (HMXB) pulsars range from a few seconds to a few hundred 
seconds and display several spin-up/down trends on time-scales ranging 
from days to years (Bildsten et al. 1997). For example, 4U 1907+09, a 
persistent supergiant X-ray binary with $P_{\mathrm{s}}$ = 440 s, is spinning down 
for more than 15 years with $\dot{P}$ =  $7.3 \times 10^{-7}$ s s$^{-1}$ 
(Baykal et al. 2001). Among LMXB pulsars, GX 1+4 has the longest spin 
period of $\sim$ 141 s. LMXB pulsars show a wide range of spin-period 
derivatives ranging from $\sim 10^{-8}$ to $\sim 10^{-11}$ s s$^{-1}$ 
(Ferrigno et al. 2007; Chakrabarty et al. 1997).
Intermediate Polars (IPs) are also slow pulsators with spin-period 
of few hundred seconds and usually show spin up phases with a typical 
spin period derivative $\sim -7 \times 10^{-11}$ s s$^{-1}$ (Patterson 1994). 
However there are a very few IPs which are found to be spinning down and 
the fastest spinning down intermediate polar PQ Gem is observed with 
$\dot{P} = 1.1 \times 10^{-10}$ s s$^{-1}$ (Mason 1997). 

The measured spin torque can be used as a clue on the nature of
the compact accretor in SAX13. The usual expression for the
accretion torque on a compact accretor is : 
$I\dot{\omega}=\dot{M}\sqrt{0.5GMr_{A}}$ (Lipunov 1992), where $I$ and 
$\dot{\omega}$ is the moment of inertia and the spin torque of the 
compact accretor respectively, $M$ and $\dot{M}$ is the mass 
and the mass accretion rate of the compact accretor respectively, 
$r_{A}$ is the Alfven radius and $G$ is the Gravitational constant. 
$r_{A}$ for a compact accretor is $\propto$ $\mu^{4/7}M^{-1/7}\dot{M}^{-2/7}$,
where $\mu$ is the magnetic moment of it. Thus the external torque
applied from the accretion disk onto the compact object scales as
$\propto M^{-3/7}R^{12/7}B^{2/7}$, where B (= $\mu R^3$) is the magnetic
field of the compact object and for a given X-ray luminosity, the value of 
$\dot{M}$ is $\propto$ $\frac{R}{M}$.
The value of $M$, $R$ and $B$ can be fixed for a typical white dwarf in 
an intermediate polar and an accreting neutron star. Thus, 
given a certain observed X-ray luminosity, the ratio of a spin torque 
for a white dwarf and a neutron star would be :  
\begin{equation}
\frac{\dot{\omega}_{wd}}{\dot{\omega}_{ns}} = 
\frac{M_{wd}^{-3/7}R_{wd}^{12/7}B_{wd}^{2/7}}{I_{wd}} \times
\frac{I_{ns}}{M_{ns}^{-3/7}R_{ns}^{12/7}B_{ns}^{2/7}}
\end{equation}

where the subscript `wd' and `ns' refer to white dwarf and neutron star
respectively.  A white dwarf has a moment of inertia $\approx 4-5$ order of
magnitude larger than a neutron star. For the typical parameters
of white dwarf and neutron star : $R_{wd} = 10^{4}$ km,
$B_{wd} = 10^{5}$ G and $M_{wd}$ = $0.1 M_{\odot}$; $R_{ns} = 10$ km, $B_{ns} = 10^{8}$ G and
$M_{ns}$ = $1.4 M_{\odot}$, the ratio between the expected accretion torques for the
two compact objects is $\approx 0.1 - 1$,  implies that we cannot 
distinguish between a white dwarf and a neutron star system on the basis 
of the spin torque. However, for a high magnetic field neutron star ($B\approx 10^{12}$ G)
the ratio between the expected accretion torques for the two compact objects is $\approx 10^{-2}$
which implies that a high B field neutron star system has a spin torque two orders of 
magnitude larger than a white dwarf. Since in SAX13, the observed spin down 
is comparable with measures of spin torques made for high B field accreting 
neutron stars, the presence of a neutron star is favored, unless the accretor 
is an intermediate polar with a high B field and a quite small mass.

The {\it XMM-Newton} spectra of SAX13 fits well with both power-law and 
black-body models. The spectral parameters of SAX13 (Table \ref{parameters}) 
are typical of an accreting neutron star X-ray pulsar with a high magnetic 
field. Accreting X-ray pulsars with low magnetic field strength usually 
display soft X-ray spectra with photon index $\ge$ 2.0 (Bildsten et al. 1997). 
IPs display hard X-ray spectra similar to high mass X-ray binary
pulsars but they also show a strong iron K$\alpha$ emission line (e.g., 
Norton et al. 1991, Muno et al. 2004). The absence of Fe K$\alpha$ 
emission line in the SAX13 X-ray spectra further makes it very unlikely 
to be an IP.

With the {\it Swift} spatial resolution, Mereghetti et al. (2008) identified 
the possible infrared counterpart of SAX13 in the 2MASS K$_S$ band image with 
magnitude 14.39 $\pm$ 0.08. Figure \ref{fig:sax13_chart} shows the ESO-NTT 
K$_S$ band image of SAX13, zoomed-in at the position of SAX13. As can be 
seen, the single 2MASS star is resolved into three stars `C1', `C2' and 
`C3' in an NTT K$_S$ band image. The position of star `C1' is consistent 
with the {\it Chandra} error circle of SAX13, and therefore `C1' is the 
most likely near-infrared counterpart of SAX13. 

Using the observed $N_{\mathrm{H}}$ = 5.8 $\times 10^{22}$ cm$^{-2}$ (Table \ref{parameters}), 
and the relation A$_V$/$N_{\mathrm{H}}$ = $5.6 \times 10^{-22}$ mag cm$^2$ (Predehl et al. 
1995), the corresponding A$_V$ for SAX13 is calculated to be 32.54 mag. Using 
the above A$_V$, we calculated the near-infrared extinction in $J$, $H$ and $K_{\mathrm{s}}$ waveband 
to be 9.01, 5.56 and 3.77 mag (Fitzpatrick 1999). Finally, the dereddened 
magnitudes of SAX13 are found to be  J = $10.56 \pm 0.11$ mag, H = $11.05 
\pm 0.09$ mag and K = $11.20 \pm 0.11$ mag. Here, we do not take into account
the uncertainties in the A$_V$. Also, we don't deny that a part of the X-ray 
absorption density could be local to the X-ray source, and not necessarily apply 
to the companion star. 

\begin{figure*}
\begin{tabular}{c}
\includegraphics[height=6cm,width=16cm]{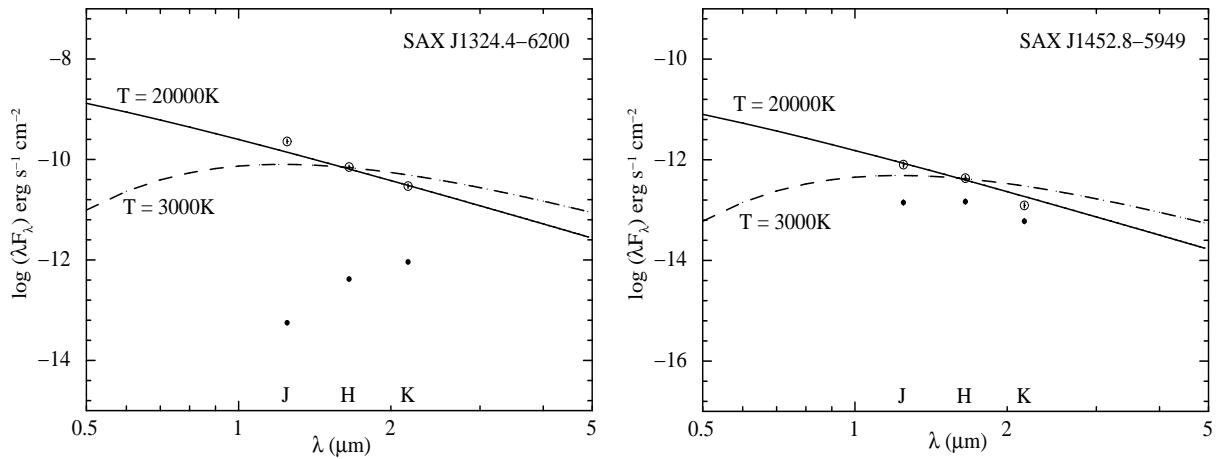}
\end{tabular}
\caption{Observed (open circles) and extinction-free (filled circles) spectral energy
densities of near-infrared counterpart of SAX1324.4--6200 ({\it{left}}) and SAX1452.8--5949 ({\it{right}}). 
The solid line and dashed-dotted line represents black-body model spectral 
energy density of the star at temperature, T = 20,000K and T = 3,000K respectively
and are shifted to match the extinction-free spectral energy density of the
star at near-infrared $H$ waveband.} 
\label{ir_s13}
\end{figure*}

With the above extinction-free {\it J, H and K$_{\mathrm{s}}$} magnitudes 
of SAX13, we used a black-body model to 
estimate the distance to different type of stars (main-sequence stars, supergiants and 
giants), for a given temperature and a radius (Cox 2000). With these calculations,  
we found that any late-type main sequence star (M through A type) would lie at a 
distance $\le$ 1.4 kpc. Thus, it is possible that the system could be an IP at a 
distance of $\le$ 1.4 kpc and in that case, the corresponding X-ray luminosity of the system would be 
$\le 10^{33}$ erg s$^{-1}$. However, it is very unlikely to find a LMXB pulsar at such a low 
X-ray luminosity. Also it is not very likely that a supergiant is the infrared counterpart 
of SAX13 as it would lie outside the Galaxy for the given magnitudes. A main sequence 
early-type star (e.g., B-type) with  temperature of 11,000 - 30,000 K and the radius of 
3.0 - 7.5 $R_{\odot}$ would 
lie at a distance $\le$ 8 kpc, indicating SAX13 could be an accreting
neutron star high mass X-ray binary pulsar. However, for the given flux densities, 
a late-type giant would also be well within the Galaxy at a similar distance.

To further confirm the above argument, we approximated the black-body model to the 
extinction-free near-infrared fluxes of SAX13 (shown as open circles in Fig \ref{ir_s13}). 
Black-body curves for temperature, T = 20,000 K and T = 3,000 K  are shown in Fig \ref{ir_s13}
as a solid and a dashed-dotted line respectively and are shifted to match the spectral
flux density of SAX13 at {\it H} wave-band. The solid line clearly indicates that SAX13 
is an early-type main-sequence star of temperature $\ge$ 20,000 K. For the sake of completeness, 
the observed fluxes of SAX13 are also shown in the same figure as filled circles.

Based on the above arguments, it is therefore, possible that SAX13 is indeed 
an accreting neutron star high mass X-ray binary pulsar at a distance of 
1.5 - 8.0 kpc. For the observed X-ray flux of SAX13 
(Table \ref{parameters}), the luminosity of the system would be 1.2 
$\times 10^{33}$ to 3.4 $\times 10^{34}$ erg s$^{-1}$. Most of the high mass X-ray pulsars 
(especially Be X-ray pulsars) are transient in nature and have been observed at luminosities 
varying from 10$^{33}$ to 10$^{38}$ erg s$^{-1}$. However SAX13 has been persistent for the past 
14 years. 
Thus it is possible that SAX13 is a persistent Be X-ray binary pulsar, 
similar to X-Persei (La Palombara et al. 2007; Reig et al. 1998). These systems might be 
part of an unusual class of accreting neutron stars with high mass companions which
have long orbital periods ($>$ 30 days) and low eccentricities (e.g., Pfahl et al. 2002).
Such long orbital period indicate that tidal circularisation cannot yet have occured
after the supernova explosion that created the neutron star. Therefore, these low eccentricities
must be primordial and the supernova explosion could not  have been accompanied
by a kick to the neutron star (Pfahl et al. 2002). 

As suggested above, a late-type giant can also be an infrared counterpart 
of SAX13, and in that case, SAX13 would be a member of 
symbiotic X-ray binary pulsars. Symbiotic X-ray binary pulsars are recently identified
as a new class of LMXBs, in which a neutron star rotates in a wide orbit around a M-type giant, 
accretes matter either from the wind of a M giant or via roche lobe overflow 
(Iben et al. 1996). Symbiotic X-ray binary pulsars are characterized by faint X-ray
emission ranging between $\sim$ 10$^{32}$ to $\sim$ 10$^{34}$ erg s$^{-1}$,  
long pulse periods and long spin-up/down phases (Masetti et al. 2007; Corbet et al. 2008). 
Thus SAX13 could also be  a symbiotic X-ray binary pulsar, at a 
distance  $\le$ 9 kpc. These systems are expected to be rare, 
owing to the short duration of both the K/M giant lifetime and the highly 
unstable mass transfer stage expected for Roche lobe overflow for LMXBs with P$_{\mathrm{orb}}$  
$\ge$ 2 days (Kalogera et al. 1996).

With the present X-ray and near-infrared observations of SAX13, we are unable to
distinguish it to be a high mass accreting X-ray pulsar or the symbiotic X-ray
binary pulsar. However, infrared spectroscopic observations would further 
help to constrain the spectral type of the infrared counterpart of 
SAX13.  

\subsection{SAX 1452.8--5949}

We detected a faint star in the ESO-NTT images
of SAX14, consistent with the {\it Chandra} error circle, shown in
Figure \ref{fig:sax13_chart}. Therefore we identify `S1' as the
infrared counterpart of SAX14. Taking $N_{\mathrm{H}}$ = $1.22 \times 10^{22}$ cm$^{-2}$
(Table \ref{parameters}), we calculated the dereddened magnitudes of SAX14
to be J = 16.66 $\pm$ 0.12 mag, H = 16.45 $\pm$ 0.12 mag and K = 17.15
$\pm$ 0.12 mag. We calculated the distance of the possible infrared
counterpart with the same method used for SAX13 (see \S 5.1) We can
rule out a supergiant, giant or an O/B-type star as the infrared
counterpart of SAX14 as, given the observed flux densities, it would
lie outside the Galaxy. This rules out the possibility that SAX14 is
a HMXB. Only a late spectral type main sequence star (M through A
type) can be in the Galaxy, at a distance $\le$ 10 kpc. A
black-body approximation to the extinction-free near-infrared fluxes of SAX14
indicate that SAX14 would have temperature $<$ 20,000 K (Fig \ref{ir_s13}). 
However, in case, a part of the X-ray absorption density
is local to the X-ray source, we would expect even lesser extinction
in near-infrared wavebands and in that case, the temperature of near-infrared
counterpart of SAX14 would be much less than 20,000 K. 
With the above arguments, we can say that SAX14 must be a binary
with a low mass companion (LMXB or IP), regardless if there are
pulsations or not.

The {\it XMM-Newton} spectrum of SAX14 is well fitted with a power-law
model of photon index 0.83 (Table \ref{parameters}). Both an accreting 
pulsar and a white dwarf are consistent with the inferred X-ray spectral 
parameters, so we cannot distinguish from the X-ray spectral information 
whether SAX14 is a neutron star or an
accreting white dwarf. Also we did not detect any strong Fe 6.4 keV line in
the X-ray spectra indicates that it is very unlikely that the system is an IP.

No significant pulsations were detected in 0.2 - 12 keV energy band for SAX14 with an
upper limit of 15\% on the fractional amplitude at 98\% confidence
level. This result is in contrast with the previous detection
made by Oosterbroek et al. (1999) who detected pulsations with a
fractional amplitude of $75\pm 25\%$.
To explain this discrepancy we are left with two possibilities: the
fractional amplitude of the pulsations decreased or the pulsations observed 
 by Oosterbroek et al. (1999) were spurious. In the later case, SAX14 can be any non pulsating
source in the Galaxy, like LMXB, accreting white dwarf, BY Dra, RS CVn or 
active star, etc.

If on the other hand the detection of pulsations by Oosterbroek et al. 
(1999) was real, then the pulse amplitude has to be reduced by a 
substantial amount. This behavior has been
observed in other LMXBs also where the pulsed fraction decreased on the 
time-scale of days (Her X-1; Ramsay et al. 2002), hours (GX 1+4; 
Naik et al. 2005) and minutes (4U 1907+09; In'T Zand et al. 1997).
Given the large value for the upper limit we obtained, the possibility 
that SAX14 is a slow pulsating LMXB or an IP is still open.

With all the above arguments, we suggest that SAX14 is not a high mass X-ray binary 
pulsar. However, the X-ray spectral, timing properties and infrared flux densities
suggest to be a low mass companion (LMXB or IP), regardless if there are
pulsations or not.


\begin{table} \label{tab:p_observation}
\centering
\caption{Spin-period history of SAX J1324.4-6200}
\label{p_observation}
\begin{tabular}{|llll|}
\hline
Telescope      & Date         & Spin period & References   \\ 
               &       (UT)        & (s)  &             \\ \hline
{\it ASCA}       & 04 Aug 1994   & 170.35  $\pm$ 0.48    & a   \\
{\it BeppoSAX}   & 22 Aug 1997   & 170.84  $\pm$ 0.04    & a   \\
{\it ASCA}       & 02 Feb 2000   & 171.25  $\pm$ 0.01   & b   \\
{\it Swift}      & 30 Dec 2007   & 172.84  $\pm$ 0.1     & c   \\
{\it XMM-Newton} & 11 Jan 2008   & 172.86  $\pm$ 0.02    & d   \\
\hline
\end{tabular}
\flushleft
REFERENCES. -- (a) Angelini et al. 1998; (b) Lin et al. 2002; (c) 
Mereghetti et al. 2008; (d) present paper.  
\end{table}


\end{document}